\documentclass[letterpaper]{article} 
\usepackage[preprint]{aaai2027}  
\nocopyright
\usepackage[hyphens]{url}  
\usepackage{graphicx} 
\usepackage{natbib}  
\usepackage{caption} 
\usepackage{makecell}
\usepackage{booktabs}
\usepackage{tabularx}
\usepackage{array}
\usepackage{dsfont}
\usepackage{algorithm}
\usepackage{algorithmic}
\usepackage[table]{xcolor}
\usepackage{amsmath,amssymb}
\usepackage{booktabs}
\newcommand{\Zhat}{\hat{Z}}
\newcommand{\present}{\textsc{present}}
\newcommand{\absent}{\textsc{absent}}
\newcommand{\invalid}{\textsc{invalidated}}
\usepackage{newfloat}
\usepackage{listings}
\DeclareCaptionStyle{ruled}{labelfont=normalfont,labelsep=colon,strut=off} 
\floatstyle{ruled}
\newfloat{listing}{tb}{lst}{}
\floatname{listing}{Listing}

\usepackage{booktabs}
\usepackage{graphicx}
\usepackage{tikz}
\title{%
\begin{tabular}{@{}c@{\hspace{0.01cm}}c@{}}
    \raisebox{-0.12\height}{%
        \includegraphics[width=1.5cm]{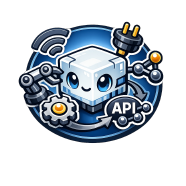}%
    }
    &
    \begin{tabular}{c}
        MADE: Belief-Driven Dual-Agent Coordination\\[-1pt]
        for Autonomous Model Deployment
    \end{tabular}
\end{tabular}%
}
\author{
    Yicheng Liu\textsuperscript{\rm 1,2},
    Bolin Zhang\textsuperscript{\rm 1,2,3}\corresponding,
    Weiran Liu\textsuperscript{\rm 1},
    Yakun Zhang\textsuperscript{\rm 4},
    Yangqin Jiang\textsuperscript{\rm 5},
    Zhiying Tu\textsuperscript{\rm 1,2},
    Dianhui Chu\textsuperscript{\rm 1,2}
}

\affiliations{
    \textsuperscript{\rm 1}Harbin Institute of Technology, Weihai, China\\
    \textsuperscript{\rm 2}Shandong Key Laboratory of Digital Service Computing
    Technology and Systems (DiSC Lab), Weihai, China\\
    \textsuperscript{\rm 3}Harbin Institute of Technology Qingdao Research Institute, Qingdao, China\\
    \textsuperscript{\rm 4}Harbin Institute of Technology, Shenzhen, China\\
    \textsuperscript{\rm 5}The University of Hong Kong, Hong Kong, China\\
    bolin@hit.edu.cn
}

\begin{document}

\maketitle

\begin{abstract}
LLM-based agents now have strong general capabilities. However, they still struggle with domain-specific tasks, motivating the integration of external tools to broaden their capabilities. The open-source community offers a vast array of AI models typically released as heterogeneous research artifacts, whereas transforming them into ready-to-call APIs is costly and labor-intensive. Automated model deployment is therefore essential for bridging the gap between model resources and tool usability, yet it remains a long-horizon, multi-stage task that has not been sufficiently explored. To tackle this challenge, we introduce \textbf{M}odel \textbf{A}utomated \textbf{D}eployment \textbf{E}ngine (\textbf{MADE}), a dual-agent coordination system. Specifically, given a model resource, MADE iteratively constructs and validates the deployment artifacts, updates its deployment belief based on execution feedback, and revisits invalid upstream artifacts until the model is successfully served as a ready-to-call API that can then be used by other agents. We further introduce \textbf{M2ABench}, a benchmark for the task of transforming \textbf{M}odels \textbf{to} ready-to-call \textbf{A}PIs. M2ABench comprises 122 real-world models with standardized test cases for evaluation. Experimental results demonstrate that MADE achieves a deployment success rate of 68.85\%, 
outperforming SWE-agent and OpenHands by 13.93 and 44.26 percentage points, respectively. Our code and dataset are publicly available at
\url{https://github.com/HITDiSC/MADE}.
\end{abstract}


\section{Introduction}

While large language models (LLMs) possess powerful capabilities, they often underperform 
in specific domains and tasks, necessitating the use of external tools 
to extend their operational scope~\citep{gorilla,hugginggpt}. The open-source community offers 
a vast array of AI models that could serve as such tools. 
However, these models must be deployed as APIs before LLMs 
can utilize them~\citep{bfcl}. Manual deployment is costly and 
labor-intensive~\cite{paleyes2022challenges, eken2025multivocal}. 
Consequently, research into automated model deployment is essential 
for transforming heterogeneous model resources into ready-to-call APIs~\cite{gundersen}.



\begin{figure*}[h]
    \centering
    \includegraphics[width=\textwidth]{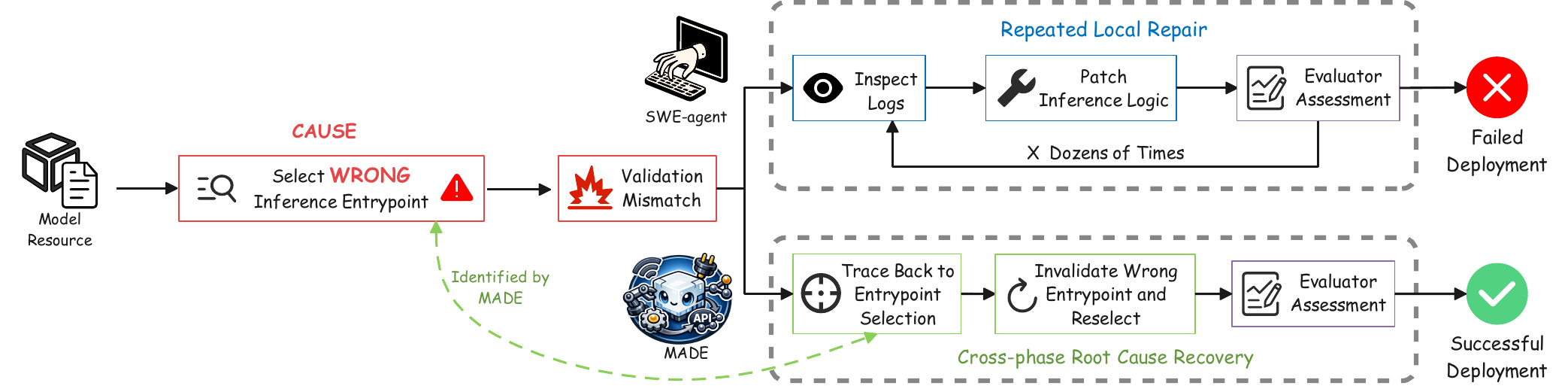}
    \caption{Illustration of delayed failure propagation and root-cause recovery. A wrong inference entry point manifests as a downstream validation mismatch, leading SWE-agent to repeatedly repair the symptom, while MADE traces the failure back to its upstream cause and corrects the erroneous selection.}
    \label{fig:motivation}
\end{figure*}

Transforming a model resource, which comprises model code, 
checkpoints, configuration files, and accompanying documentation, 
into an external tool for LLMs is not merely a matter of 
executing the provided code, but of integrating these artifacts into a ready-to-call API.
Their tight interdependence gives rise to two key challenges. 
First, deployment failures often surface far from the artifact that caused them, 
so the point of failure is not necessarily the point that should be repaired. 
As illustrated in Figure~\ref{fig:motivation}, a coding agent makes 
dozens of attempts to address a downstream error without revisiting 
the upstream artifact that caused it. 
Second, steering a deployment requires a compact and current global view of its state. 
Concrete deployment actions, however, must be grounded in firsthand execution evidence, 
which is often extensive and rapidly becomes stale.
Over a long deployment, maintaining a global state view 
while retaining the firsthand execution evidence becomes 
increasingly difficult within a single decision context.

Existing related autonomous agents are built for similar goals but stop
short at different points. Environment-building tools reconstruct 
dependencies and verify that the code runs~\citep{repo2run}. 
Building on this, reproduction and task-solving systems execute the model itself,
either to reproduce reported results~\citep{scireplicate,paperbench} 
or to generate specified outputs~\citep{repomaster,gittaskbench}.
However, these executions remain one-off and task-bound. 
They do not expose the model as a persistent ready-to-call API that agents 
can independently invoke on new inputs and receive well-formed 
responses. General-purpose coding agents such as SWE-agent~\citep{sweagent} and
OpenHands~\citep{openhands} offer greater flexibility. They are free to modify code and execute
commands. However, their broad trial-and-error exploration incurs substantial token costs, 
while still achieving limited deployment success.

To this end, we introduce \textbf{M}odel \textbf{A}utomated
\textbf{D}eployment \textbf{E}ngine (\textbf{MADE}), an LLM-based dual-agent
system that transforms a model resource into a ready-to-call API. The key
idea is that if agents can maintain a lightweight, deterministic, and
revisable belief over valid deployment artifacts, coordination can be driven by state changes
rather than repeated LLM inspection. This enables effective and cost-efficient
automated model deployment over long execution trajectories.

MADE addresses cross-stage failure attribution while coordinating 
global deployment reasoning with firsthand execution evidence. 
When a downstream failure provides evidence against an upstream artifact, 
MADE revises the belief non-monotonically, invalidating the artifact and 
redirecting execution to the stage responsible for rebuilding it. 
To combine the two complementary views, a manager uses the shared belief 
to determine which operation is appropriate in the global deployment process, 
while an executor interacts with the environment to determine whether 
that operation is locally feasible. When direct execution evidence 
conflicts with the manager's instruction, the executor can challenge or reject it, 
feeding the evidence back into the shared belief.

To enable systematic evaluation of autonomous model deployment, we introduce
a benchmark for the task of transforming \textbf{M}odels \textbf{to} ready-to-call \textbf{A}PIs, \textbf{M2ABench}. 
To our knowledge, \textbf{M2ABench} is the first benchmark for this task. 
It comprises 122 real-world open-source models spanning text, image, audio,
scientific structures, and multimodal tasks, with five human-authored test
cases for each model.
Prior work provides no end-to-end criterion for successful deployment. We
therefore define success at the deployment level: the endpoint must return
responses that conform to its declared contract and match the output structure
of the model, while using the model's own checkpoint and
inference logic rather than a substitute. Predictive accuracy is
excluded because it evaluates the model itself, not its deployment.

In summary, this work makes the following contributions:
\begin{itemize}
\item We introduce \textbf{M2ABench}, the first benchmark for transforming models to APIs, comprising 122 real-world open-source model resources across diverse modalities. We further define an end-to-end evaluation metric based on contract-conformant deployment outputs.

\item We propose \textbf{MADE}, a dual-agent system that transforms heterogeneous model resources into ready-to-call inference APIs. It coordinates long-horizon deployment through a lightweight, deterministic, and revisable belief over deployment artifacts.

\item We conduct extensive experiments on M2ABench. MADE successfully deploys 68.85\% of the 122 model resources, outperforming SWE-agent and OpenHands by 13.93 and 44.26 percentage points, respectively.
\end{itemize}
  
We envision automated model deployment as foundational infrastructure 
for future AI agents, with MADE providing a concrete step towards 
making diverse specialized model capabilities directly callable.

\section{Related Work}
\paragraph{Repository-to-runnable and reproduction.}
Prior work automates parts of the path from source to execution but stops
before a callable service. Systems build environments and navigate
repositories~\citep{repo2run,r2e,repomaster}, while benchmarks score
compilation and setup~\citep{envbench}, code against reference
implementations~\citep{scireplicate}, or a single execution of a reported
experiment or task~\citep{paperbench,gittaskbench,artifactdeploy}. A separate
line supplies tools by \emph{writing code}~\citep{voyager,creator,gorilla}.
\textbf{We differ:} We provide a \emph{served},
contract-conformant API exercised end-to-end, and we deploy pretrained
research \emph{models}, whose capabilities cannot be hand-written.

\paragraph{Coordination and supervision in multi-agent systems.}
Existing multi-agent systems with manager--worker architectures coordinate on fixed
structure~\citep{metagpt,autogen} or intervene at round
boundaries~\citep{agentverse}, while uncertainty-based supervisors defer on
local signals~\citep{knowno}. Disagreement is resolved by
majority~\citep{mad}, which can suppress a correct minority~\citep{madstudy},
and principal--agent analyses treat an executor's private information as a
hazard to suppress~\citep{principalagent}. Shared state is repaired after the
fact, by compensating rollback~\citep{sagallm} or post-hoc
attribution~\citep{causalflow,dover}; non-monotonic revision itself is
classical~\citep{tms}. \textbf{We differ}: MADE intervenes on
\emph{inconsistency} between a proposed action and a global artifact state,
invalidates upstream artifacts \emph{live}, and \emph{grants} the
tool-grounded executor authority to override the manager.

\section{Preliminaries}
\subsection{Task Formulation}
Given a model resource $R$ and a deployment host $H$ with available compute resources and network access, we define autonomous model deployment as the task of constructing a complete deployment artifact $\mathcal{D}$:
\begin{equation}
\mathcal{D} =
\big(
E,\ W_\mathcal{R},\ f_\mathcal{D},\ \sigma_{\mathrm{in}},
\sigma_{\mathrm{out}},\ \mathcal{A}
\big),
\label{eq}
\end{equation}
where $E$ is an execution environment where the model code runs; $W_\mathcal{R}$ denotes the checkpoints of the model; $f_\mathcal{D}$ is a single-instance inference procedure implementing the preprocessing, model execution, and postprocessing; $(\sigma_{\mathrm{in}},\sigma_{\mathrm{out}})$ is an explicit, machine-readable input/output contract; and $\mathcal{A}$ is a served API that exposes $f_\mathcal{D}$ in $E$.

Training, fine-tuning, batching, streaming, and production-scale serving
optimizations are outside the scope of this task.

\subsection{M2ABench}
\begin{table}[t]
\centering
\label{tab:Bench}
\renewcommand{\arraystretch}{1.12}
\small
\begin{tabularx}{\columnwidth}{
    >{\centering\arraybackslash}X
    >{\centering\arraybackslash}X
}
\toprule
\textbf{Category} & \textbf{\# Models (\%)} \\
\midrule
Text                 & 49 (40.16\%) \\
Audio                & 13 (10.66\%)   \\
Image                & 27 (22.13\%) \\
Scientific Structure & 25 (20.49\%) \\
Multimodal           & 8 (6.56\%)   \\
\midrule

\textbf{Total}
& \textbf{122} \\
\bottomrule
\end{tabularx}
\caption{Input Distribution of M2ABench}
\label{tab:Bench}
\end{table}






We propose \textbf{M2ABench}, a benchmark containing 122 real-world 
model resources across five modalities. Each model is paired with 
a test suite $T_\mathcal{R}={(x_j,y_j)}_{j=1}^{5}$ consisting of five human-authored cases. 
The input $x_j$ and output $y_j$ examples are derived from materials provided by the model, 
including documented usage examples and samples from its stated training datasets. 
Although the models cover diverse modalities and tasks, 
their evaluated outputs are all represented in textual form to 
enable unified comparison. We manually checked that all resources 
required for deployment are in place. The models are collected from papers published 
at top-tier venues and well-known open-source projects. We use them in 
their original form without modifying their code or documentation. 
As a result, M2ABench evaluates whether a deployment system 
can handle diverse runtime environments, input formats, and inference procedures.

Table~\ref{tab:Bench} illustrates the input distribution of M2ABench. The \emph{scientific structure} includes domain-specific structured data, such as proteins, molecules, and physiological signals.

\subsection{Evaluation Metrics} 
\paragraph{Success Definition.}
We consider a deployment successful if its endpoint produces a structurally consistent output for every input of the model resource. Let $\mathrm{S}(z)$ denote the structure of an output $z$. Formally, we define the successful deployment as:
\begin{equation}
\textsc{Success}(\mathcal{D};\mathcal{R})
\iff
\mathrm{S}\bigl(A(X_\mathcal{R})\bigr)
\cong
\mathrm{S}^{}(\mathcal{R}),
\label{eq}
\end{equation}
where $X_R$ is the input space, $\mathrm{S}^{}(\mathcal{R})$ is the output structure and $A(X_\mathcal{R})$ denotes the API outputs produced for $X_\mathcal{R}$. Structural consistency requires the returned output to have the expected types, nesting, and formats. Since $\mathrm{S}\bigl(A(X_R)\bigr)$ is defined only when the endpoint successfully returns a valid response, any execution failure violates this condition.

This definition evaluates deployment correctness rather than model capability. It does not assess predictive accuracy, which depends on the underlying model and cannot be defined uniformly across heterogeneous tasks.

\paragraph{Deployment Success Rate (DSR).}
A deployment is considered successful only if the endpoint output has the same structure as the corresponding expected output for all five test cases.
For each test case $(x_i,y_i)\in T_R$, define
\begin{equation}
P_i
=
\prod_{j=1}^{5}\mathds{1}\!\left[
\mathrm{S}\bigl(\mathcal{A}(x_j)\bigr)
\cong
\mathrm{S}(y_j)
\right].
\label{eq:test-pass}
\end{equation}

\begin{equation}
\mathrm{DSR}
=
\frac{1}{N}
\sum_{i=1}^{N}
P_i
\label{eq:dsr}
\end{equation}

A deployment is counted as unsuccessful if any test case fails to produce a valid response, its output does not match the corresponding expected output structure, or no deployment artifact is produced.
\paragraph{LLM Calls.} The number of LLM invocations issued while deploying a resource, reported per successful deployment. 
\paragraph{Token Consumption.} The total prompt and completion tokens consumed across LLM calls.


\section{Methodology}

\begin{figure*}[t]
    \centering
    \includegraphics[width=\textwidth]{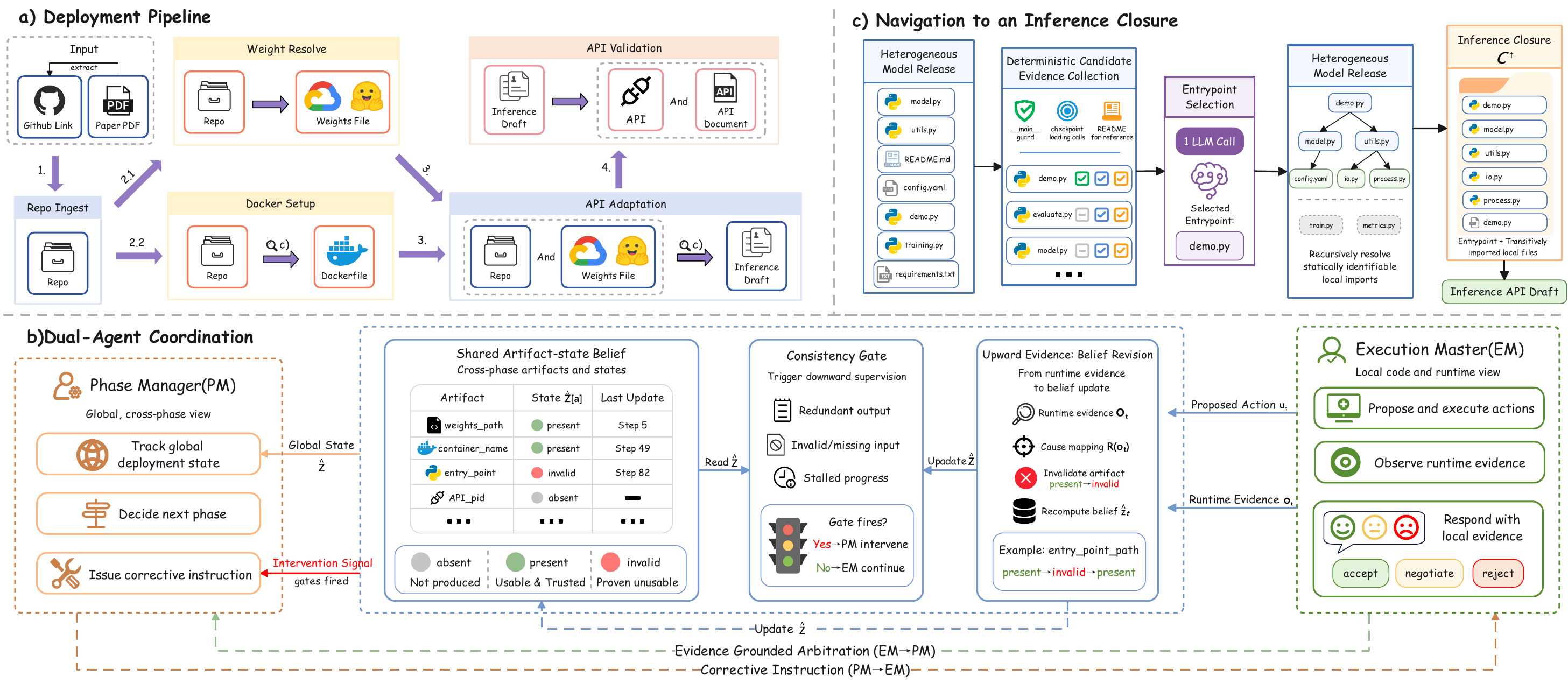}
    \caption{Overview of MADE consisting of three main components: a) Deployment Pipeline, b) Dual-Agent Coordination, and c) Navigation to an Inference Closure.}
    \label{fig:overview}
\end{figure*}

As illustrated in Figure~\ref{fig:overview}, MADE is a dual-agent coordination system 
that executes the automated deployment. Within each
phase in Table~\ref{tab:pipeline}, an \emph{Execution Master} (EM) selects tools and gathers firsthand
evidence about the code and execution environment. Across phases, a
\emph{Phase Manager} (PM) determines which phase should run next and whether
its outputs are valid. The two agents coordinate through a shared artifact store and an explicit
belief over the deployment state. An artifact may appear valid until a
dependent component exercises it, allowing failures observed in later
phases to retroactively invalidate artifacts produced earlier.

\subsection{Deployment Pipeline}
The five phases (in Figure~\ref{fig:overview}a) follow the dependency order of deployment artifacts and
jointly construct $D$. Faithfulness is enforced by design rather than
assumed. WeightResolve retrieves $W_R$ only from model-specified sources,
and APIAdaptation reconstructs $f_D$ from the model's own inference
code. The resulting inference environment and serving code are executed
inside isolated Docker containers to ensure dependency isolation,
reproducibility, and safe execution without affecting the host system.
Table~\ref{tab:pipeline} summarizes the full pipeline. Below, we introduce
three key mechanisms that combine LLM generation with deterministic checks
and execution feedback to detect and correct deployment failures.

\begin{table*}[t]
\centering
\small

\renewcommand{\arraystretch}{1.08}
\renewcommand{\tabularxcolumn}[1]{m{#1}}
\tabcolsep=5pt

\begin{tabularx}{\textwidth}{
    @{}
    >{\raggedright\arraybackslash}m{0.13\textwidth}
    >{\raggedright\arraybackslash}m{0.19\textwidth}
    >{\raggedright\arraybackslash}X
    >{\raggedright\arraybackslash}m{0.23\textwidth}
    @{}
}
\toprule
\textbf{Phase} & \textbf{Input} & \textbf{Function} & \textbf{Output} \\
\midrule

RepoIngest
& model resource $R$
& Parses the resource and identifies the model's task.
& structured resource: repository layout, task type \\

DockerSetup
& structured resource; \newline host GPU profile
& Builds the execution environment.
& $E$: a running container \\

WeightResolve
& structured resource
& Locates and retrieves the model checkpoints.
& $W_R$: locally available checkpoints \\

APIAdaptation
& structured resource; \newline running container; \newline resolved checkpoints
& Selects the code needed for inference, reconstructs the
inference procedure, and exposes it as a service with its I/O contract.
& $f_D,\ (\sigma_{\mathrm{in}},\sigma_{\mathrm{out}})$:
service code and its declared contract \\
\addlinespace[3pt]

APIValidation
& service code; \newline declared contract; \newline running container
& Serves the endpoint, validates it against the success criterion, and
repairs failures.
& $A$: a ready-to-call API \\

\bottomrule
\end{tabularx}
\caption{The MADE pipeline. Together, the phases deliver the components of
the target artifact $D$. After RepoIngest, DockerSetup and
WeightResolve proceed independently and in parallel.}
\label{tab:pipeline}
\end{table*}

\paragraph{Execution-guided environment correction.}
DockerSetup reconstructs the execution environment by
combining model's declared inference dependencies with dependencies inferred from
the code. It then generates a containerized environment tailored to the host
GPU profile. Environment construction is performed through an execution-guided
correction loop: failures during image building or inference execution provide
concrete error evidence that is fed back to revise the environment
configuration. This allows MADE to recover from environment mismatches, such as
adapting the PyTorch installation strategy after repeated CUDA compatibility
failures.

\paragraph{Inference--contract co-generation.}
APIAdaptation reconstructs $f_D$ from the model's inference code,
covering model loading, preprocessing, model execution, and postprocessing.
The generated implementation is grounded in the constructed environment
and resolved checkpoints, ensuring that it matches the dependencies and
artifacts available at runtime. Rather than generating the API contract
separately, MADE derives $\sigma_{\mathrm{in}}$ and
$\sigma_{\mathrm{out}}$ alongside the corresponding preprocessing and
postprocessing logic. It then checks that the fields declared by
$\sigma_{\mathrm{in}}$ match those accessed by the generated
preprocessing code. The inference procedure and its API contract are
therefore constructed and validated as a single consistent artifact.

\paragraph{Criterion-guided validation and bounded repair.}
APIValidation launches the service, verifies its health, 
and invokes the M2ABench evaluator to test the live endpoint. 
The evaluator constructs requests according to the declared input 
contract and checks whether the responses satisfy the success criterion. 
When validation fails, APIValidation locally repairs minor issues 
that can be resolved without changing artifacts produced by earlier phases, 
such as missing auxiliary files, ABI-incompatible dependencies, 
and isolated service-code defects. More substantial failures that 
indicate problems in upstream artifacts, such as an incorrect environment, 
checkpoint, or inference procedure, are reported to 
the coordination layer for cross-phase resolution.

\subsection{Dual-Agent Coordination}

As shown in Figure~\ref{fig:overview}b, MADE separates cross-phase coordination from in-phase execution 
between two complementary agents. PM tracks the global deployment state 
and determines what should be done next, while EM interacts with the code 
and runtime environment to determine what can actually be done. 
Rather than allowing either agent to act alone, 
MADE coordinates their decisions through a shared, revisable belief, 
while allowing firsthand execution evidence to correct outdated global judgments.

\paragraph{Artifact-state belief.}
MADE maintains a global belief about the artifacts produced and consumed across deployment phases. This belief records whether each artifact is currently available and whether it can still be trusted by downstream phases. Let $\mathcal{A}$ denote the set of cross-phase artifacts listed in Table~\ref{tab:pipeline}. For each artifact $a\in\mathcal{A}$, the belief $\Zhat[a]$ takes one of three states: 
\begin{itemize}
    \item $\absent$: the artifact has not been produced;
    \item $\present$: the artifact exists and has not been invalidated;
    \item $\invalid$: later execution evidence has shown the artifact to be unusable.
\end{itemize}
The belief is maintained automatically without LLM calls.

\paragraph{Upward evidence: non-monotonic invalidation.}
A failure observed in a later phase may reveal that an earlier artifact
is faulty. MADE records such artifacts in an invalidation set $I$ and
updates the belief accordingly. Given failure evidence $o_t$, let
$\mathcal{R}(o_t)\subseteq\mathcal{A}$ denote the set of earlier artifacts
identified as candidate root causes of the failure. MADE updates
\begin{equation}
I_{t+1}=I_t\cup\mathcal{R}(o_t).
\label{eq:update}
\end{equation}
The belief is then recomputed, changing the implicated artifacts from
$\present$ to $\invalid$. Once an artifact is successfully regenerated,
it is removed from $I$. To avoid unnecessary backtracking,
$\mathcal{R}$ considers only artifacts produced by previously executed
phases, applies conservative keyword rules, and requires PM confirmation.

\paragraph{Downward control: consistency-gated supervision.}
Continuous auditing is costly, whereas boundary-only review allows errors
to compound. MADE therefore intervenes only when EM's proposed action
$u_t$, with declared inputs $\textsc{In}(u_t)$ and outputs
$\textsc{Out}(u_t)$, conflicts with $\Zhat$, or the recent in-phase
history $h_t$ indicates stalled progress:
\begin{equation}
\begin{aligned}
&\hspace{-3em}\textsc{Intervene}(u_t,h_t;\Zhat)
\\[-2pt]
&=
\begin{cases}
1, &
\exists a\in\textsc{Out}(u_t):
\Zhat[a]=\present,
\\
1, &
\exists a\in\textsc{In}(u_t):
\Zhat[a]\neq\present,
\\
1, &
h_t\text{ indicates stalled progress},
\\
0, &
\text{otherwise}.
\end{cases}
\end{aligned}
\label{eq:gate}
\end{equation}
Stalled progress occurs when the same tool is used $k$ times without a
new artifact or when at least $m$ consecutive turns end in errors.

The invalid-input condition couples the two directions: once an execution
failure invalidates an artifact (a CUDA error demoting the container, say),
the gate fires on proposals that still depend on it. Gate firing is
deterministic; only after a fire does PM invoke one LLM step to generate
the corrective instruction. A per-phase cap and cooldown keep
interventions sparse, while a policy switch (never, boundary-only, always,
random) supports the supervision-policy ablation in the experiments.

\paragraph{Conflict arbitration: evidence-grounded override.}
Because $\Zhat$ may lag behind the current execution state, 
PM instructions are not unconditionally binding. EM responds with
$d_t\in{\textsc{accept},\textsc{negotiate},\textsc{reject}}$ and, when
challenging an instruction, provides supporting local evidence. Negotiation
raises a concrete objection for bounded reconsideration by PM, whereas
rejection allows EM to proceed with its own plan while reporting the reason.
This mechanism ensures that firsthand execution evidence can override an
instruction derived from a stale global belief.

\paragraph{Coupling and cost.}
These mechanisms form a closed loop. Execution evidence invalidates
artifacts and updates the belief; the gate reads the updated belief to
constrain EM's proposals; override arbitrates conflicts between the
resulting PM instruction and EM's local evidence; and the next execution
outcome triggers another belief update. Belief computation and gate
decisions are deterministic, while LLM-based supervision is invoked only
when an inconsistency or stall is detected. PM supervision therefore
scales with the number of such signals rather than with the full execution trajectory. 

\subsection{Navigation to an Inference Closure}
As illustrated in Figure~\ref{fig:overview}c, MADE navigates the repository to identify an inference closure.
Reconstructing $f_D$ requires the model's inference code, 
which is often scattered across multiple scripts and intertwined with 
training and evaluation logic. File selection in this setting 
has an asymmetric failure mode. Omitting even one configuration loader or 
utility module may cause reconstruction to fail entirely, 
whereas including irrelevant files mainly consumes the limited context budget. 
MADE therefore does not rank repository files by relevance. 
Instead, it identifies an \emph{inference closure} $C^{\dagger}$, 
consisting of the inference entry point and all local repository files that it transitively imports.

MADE confines the only semantic decision in this process to a single LLM call. 
A deterministic pass first collects deployment-specific evidence 
for each candidate entry file. This evidence includes a \texttt{\_\_main\_\_} guard, 
AST-detected checkpoint-loading calls, and references to 
the file or its invocation command in the distilled README. 
A single LLM call then resolves the remaining ambiguity 
among the candidates using this evidence and the task identified during ingestion.

Once the entry point is selected, MADE constructs $C^{\dagger}$ 
deterministically by recursively resolving statically 
identifiable local imports. APIAdaptation uses the resulting files 
as its repository-code context, reducing irrelevant code 
while preserving the local dependencies exposed by the static import graph. 
Closure construction therefore requires neither additional relevance ranking nor further LLM calls.

\section{Experiments}

\subsection{Setup}
\paragraph{Basic Settings}
For all methods, we use GPT-5.2 as the backbone LLM with a sampling temperature of 0.2. MADE and all baselines are evaluated on the same local server under identical hardware and runtime constraints. Each deployment run is allocated two NVIDIA L20 GPUs. Repositories requiring GPU acceleration are granted access to both GPUs during environment construction, inference execution, and API validation. In addition, all methods are evaluated using the same evaluator provided by M2ABench, ensuring a consistent and fair evaluation protocol.

\paragraph{Baselines}
We compare MADE with two representative software-engineering agents with complementary designs: SWE-agent~\cite{sweagent}, a task-oriented coding agent developed for resolving repository-level issues, and OpenHands~\cite{openhands}, a general-purpose agent for autonomous software development. Each method receives the same model resource, deployment instruction, hardware and runtime resources, and access to the same validation interface. All methods may use the evaluator's returned results to iteratively revise their deployments. The evaluator and its validation cases are isolated from the agents and cannot be inspected or modified. Deployment success is determined using the same criterion and evaluator.

\textbf{SWE-agent.}
SWE-agent resolves real-world software issues through repository navigation, code modification, command execution, and iterative debugging.

\textbf{OpenHands.}
OpenHands is a software-development agent that can inspect repositories, modify files, execute commands, and interact with development environments.

\subsection{Main Results}
\begin{table}[t]
\centering

\small
\tabcolsep=4pt
\renewcommand{\arraystretch}{1.08}

\begin{tabularx}{\columnwidth}{
    @{}
    l
    c
    >{\centering\arraybackslash}X
    >{\centering\arraybackslash}X
    @{}
}
\toprule
\rowcolor{gray!20}
\textbf{Method}
& \textbf{DSR}
& \makecell[c]{\textbf{Avg. LLM Calls}\\\textbf{per Success}}
& \makecell[c]{\textbf{Avg. Tokens}\\\textbf{per Success}} \\
\midrule

SWE-agent
& 54.92\%~(67)
& 89.76
& 3379.480k \\

OpenHands
& 24.59\%~(30)
& 21.07
& 846.117k \\

\midrule

\textbf{MADE}
& \textbf{68.85\%~(84)}
& \textbf{129.81}
& \textbf{1058.297k} \\

\bottomrule
\end{tabularx}

\caption{Comparison of baselines in deployment success rate (DSR), number of successful deployments, and average LLM calls and tokens per successful deployment.}
\label{tab:baseline_results}
\end{table}
Table~\ref{tab:baseline_results} reports the deployment effectiveness and inference cost of MADE and the two agent-based baselines. MADE achieves the highest DSR of 68.85\%, successfully deploying 84 of the 122 model resources. This exceeds SWE-agent by 13.93 percentage points and OpenHands by 44.26 percentage points. The substantial margin over both baselines demonstrates that a deployment-specific workflow is more reliable than directly adapting general-purpose software-engineering agents to heterogeneous model resources.

MADE also achieves a favorable balance between deployment success and token consumption. It requires an average of 1058.297k tokens per successful deployment, reducing token usage by 68.68\% compared with SWE-agent while attaining a considerably higher DSR. OpenHands consumes fewer tokens per successful deployment than MADE, at 846.117k tokens, but achieves a DSR of only 24.59\%, indicating that low token consumption alone does not translate into reliable deployment.

Notably, MADE makes more LLM calls per successful deployment than either baseline, but each call processes a substantially smaller amount of context on average than those of SWE-agent and OpenHands. This pattern is consistent with MADE's design, which decomposes deployment into frequent, scope-limited decisions supported by deterministic navigation and validation, rather than relying on a smaller number of expensive long-context interactions. 

Overall, MADE achieves a balance between deployment effectiveness and efficiency.

\subsection{Token Consumption Analysis}
\begin{figure}[t]
    \centering
    \includegraphics[width=\linewidth]{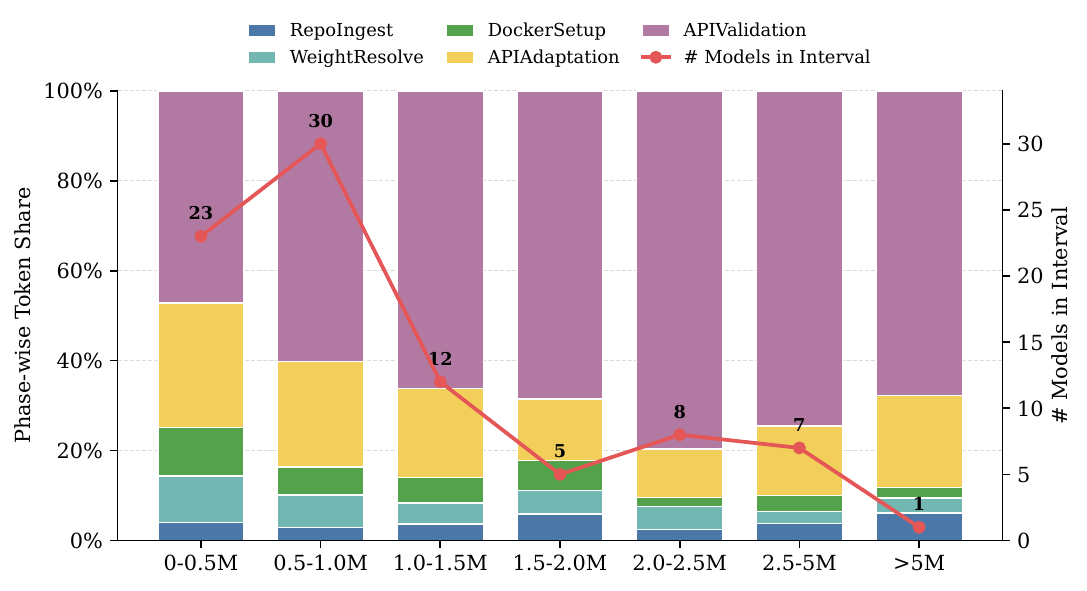}
    \caption{Phase-wise token consumption distribution across different total-token intervals. The stacked bars show the relative token proportion of each phase, and the red line indicates the number of models in each interval.}
    \label{fig:token_consumption}
\end{figure}
Figure~\ref{fig:token_consumption} groups deployment instances by their total token consumption and reports the relative contribution of each phase within every interval. Most instances consume fewer than 1.5M tokens, while the number of instances decreases substantially in higher intervals. Only a small number exceed 2.5M tokens, revealing a long-tail distribution in which a few difficult repositories account for disproportionately high costs. The interval above 5M tokens should be interpreted cautiously because it contains very few instances.

The phase composition shows that token consumption is concentrated mainly in APIValidation and APIAdaptation. APIValidation accounts for the largest share across all intervals, and its proportion generally increases with total consumption. This indicates that high-cost deployments are primarily driven by repeated cycles of execution, failure diagnosis, repair, and revalidation. APIAdaptation also contributes a substantial share because reconstructing a callable inference procedure requires identifying and integrating fragmented preprocessing, model initialization, checkpoint loading, execution, and postprocessing logic. These components are often distributed across multiple files and intertwined with training or evaluation code, making inference reconstruction reasoning intensive. In contrast, RepoIngest, DockerSetup, and WeightResolve contribute relatively small proportions.


\subsection{Ablation Study}
\begin{table}[t]
\centering

\renewcommand{\arraystretch}{1.12}
\tabcolsep=3pt
\renewcommand{\tabularxcolumn}[1]{m{#1}}

\small
\begin{tabularx}{\columnwidth}{
    @{}
    >{\raggedright\arraybackslash}m{2.8cm}
    >{\centering\arraybackslash}m{0.85cm}
    @{\hspace{20pt}}
    >{\centering\arraybackslash}m{1.15cm}
    >{\centering\arraybackslash}X
    @{}
}
\toprule
\multicolumn{1}{c}{\textbf{Variant}}
& \makecell{\textbf{DSR}\\\textbf{(\%)}}
& \makecell{\boldmath$\Delta$\textbf{DSR}\\\textbf{(pp)}}
& \makecell{\textbf{\# Successful}\\\textbf{Deployments}} \\
\midrule

w/o Dual-Agent Coordination
& 27.87
& $-40.98$
& 34 \\

w/o Navigation to an Inference Closure
& 16.39
& $-52.46$
& 20 \\

\midrule
\multicolumn{1}{c}{\textbf{MADE}}
& \textbf{68.85}
& \textbf{--}
& \textbf{84} \\
\bottomrule
\end{tabularx}
\caption{Ablation results on M2ABench. $\Delta$DSR denotes the
absolute decrease from the full MADE in percentage points.}
\label{tab:ablation_results}
\end{table}
To evaluate the contributions of inference-closure navigation and dual-agent coordination, we construct two ablated variants of MADE by removing each component separately. Without dual-agent coordination, the Phase Manager and Execution Master are replaced by a single agent while the remaining deployment pipeline and tools are preserved. Without navigation to an inference closure, the agent no longer identifies an inference entry point and recursively resolves its local dependencies, instead relying on generic repository inspection. Table~\ref{tab:ablation_results} reports the results.

Removing dual-agent coordination decreases DSR from 68.85\% to 27.87\%, with the number of successful deployments falling from 84 to 34. This substantial degradation demonstrates the importance of coordinating cross-phase artifact state with firsthand execution evidence. Without this mechanism, a single agent is less effective at revising earlier deployment decisions when failures surface in later phases.

Removing navigation to an inference closure causes an even larger DSR decrease of 52.46 percentage points from 68.85\%, and reduces the number of successful deployments from 84 to 20. This result indicates that generic file inspection is insufficient for reliably reconstructing inference procedures from heterogeneous research repositories. Identifying the inference entry point and deterministically expanding its local dependencies are therefore critical for supplying API adaptation with the complete inference-related code context.

Overall, both components contribute substantially to MADE's deployment performance. Inference-closure navigation provides the repository context required to reconstruct a runnable inference procedure, whereas dual-agent coordination enables failures observed during execution to revise and redirect decisions across deployment phases.

\subsection{Failure Case Study}
\begin{table}[t]
\centering

\renewcommand{\arraystretch}{1.08}
\renewcommand{\tabularxcolumn}[1]{m{#1}}
\tabcolsep=3pt

\small
\begin{tabularx}{\columnwidth}{
    @{}
    >{\raggedright\arraybackslash}m{3cm}
    >{\raggedright\arraybackslash}X
    >{\centering\arraybackslash}m{1.6cm}
    @{}
}
\toprule
\rowcolor{gray!20}
\multicolumn{1}{c}{\textbf{Category}}
& \multicolumn{1}{c}{\textbf{Description}}
& \multicolumn{1}{c}{\textbf{\# Cases (\%)}} \\
\midrule

Service Startup
& API Service startup or request-handling failure
& 4 (10.53\%) \\
\addlinespace[3pt]

Contract Mismatch
& Input or output contract violation
& 4 (10.53\%) \\
\addlinespace[3pt]

CUDA Runtime
& CUDA-related model execution failure
& 2 (5.26\%) \\
\addlinespace[3pt]

Container Build
& Docker container build failure
& 4 (10.53\%) \\
\addlinespace[3pt]

Import Resolution
& Missing package or repository module
& 5 (13.16\%) \\
\addlinespace[3pt]

Checkpoint Resolution
& Checkpoint location or acquisition failure
& 19 (50.00\%) \\

\midrule
\multicolumn{2}{l}{\textbf{Total}}
& \textbf{38 (100\%)} \\
\bottomrule
\end{tabularx}
\caption{Breakdown of failure categories among unsuccessful MADE deployments.}
\label{tab:failed_cases}
\end{table}
To understand the limitations of MADE, we manually inspect the execution logs and intermediate artifacts of all unsuccessful deployments and assign each case to the first root cause that prevents the pipeline from proceeding. As shown in Table~\ref{tab:failed_cases}, missing or unresolvable model weights account for 50\% of all failed cases. This proportion should not be interpreted as evidence that the remaining deployment steps would necessarily succeed. Rather, these deployments terminate during \texttt{WeightResolve}, before API adaptation, service construction, and endpoint validation can be exercised. The result therefore highlights model-resource availability as the earliest and most frequently observed bottleneck in the current pipeline.

Most of the remaining failures are related to environment reconstruction and runtime compatibility. Import failures arise from missing packages, undeclared repository modules, or inconsistent dependency specifications, while Docker, CUDA, and port-related failures reflect incompatibilities between the released code and the target execution environment. Together, these cases show that reproducing the runtime assumptions of heterogeneous research repositories remains difficult even when the required model resources can be obtained.

A smaller number of failures occur during service construction and interface validation, including service-startup errors, API-construction errors, and input/output contract mismatches. These results suggest that MADE's remaining limitations span multiple stages, but the failures observed most frequently arise before or during execution-environment preparation rather than from standardized API construction alone.

\section{Conclusion}
In this work, we formulate autonomous model deployment as the task of transforming heterogeneous model resources into ready-to-call APIs, and introduce M2ABench for systematic evaluation. We propose MADE, a belief-driven dual-agent system that combines inference-closure navigation with cross-phase artifact tracking and evidence-grounded recovery. Experiments show that MADE substantially outperforms general-purpose coding agents, while ablation studies confirm the importance of both code navigation and dual-agent coordination. We hope this work provides a foundation for making open-source models more accessible as reliable tools for LLM agents.

\section{Acknowledgment}
This work was supported by the Shandong Provincial Natural Science Foundation under Grant No. ZR2026QC1033, the Special Funding Program for Taishan Scholars of Shandong Province and the National Natural Science Foundation of China under Grant No. 62472121.

\bibliography{made}

@inproceedings{openhands, title={Openhands: An open platform for ai software developers as generalist agents}, author={Wang, Xingyao and Li, Boxuan and Song, Yufan and Xu, Frank F and Tang, Xiangru and Zhuge, Mingchen and Pan, Jiayi and Song, Yueqi and Li, Bowen and Singh, Jaskirat and others}, booktitle={International Conference on Learning Representations}, volume={2025}, pages={65882--65919}, year={2025}}

@article{sweagent, title={Swe-agent: Agent-computer interfaces enable automated software engineering}, author={Yang, John and Jimenez, Carlos and Wettig, Alexander and Lieret, Kilian and Yao, Shunyu and Narasimhan, Karthik and Press, Ofir}, journal={Advances in Neural Information Processing Systems}, volume={37}, pages={50528--50652}, year={2024}}

@article{voyager, title={Voyager: An open-ended embodied agent with large language models}, author={Wang, Guanzhi and Xie, Yuqi and Jiang, Yunfan and Mandlekar, Ajay and Xiao, Chaowei and Zhu, Yuke and Fan, Linxi and Anandkumar, Anima}, journal={arXiv preprint arXiv:2305.16291}, year={2023}}

@article{hugginggpt, title={Hugginggpt: Solving ai tasks with chatgpt and its friends in hugging face}, author={Shen, Yongliang and Song, Kaitao and Tan, Xu and Li, Dongsheng and Lu, Weiming and Zhuang, Yueting}, journal={Advances in Neural Information Processing Systems}, volume={36}, pages={38154--38180}, year={2023}}

@article{gorilla, title={Gorilla: Large language model connected with massive apis},
  author={Patil, Shishir G and Zhang, Tianjun and Wang, Xin and Gonzalez, Joseph E},
  journal={Advances in Neural Information Processing Systems},
  volume={37},
  pages={126544--126565},
  year={2024}}

@inproceedings{gundersen, title={State of the art: Reproducibility in artificial intelligence}, author={Gundersen, Odd Erik and Kjensmo, Sigbj{\o}rn}, booktitle={Proceedings of the AAAI conference on artificial intelligence}, volume={32}, number={1}, year={2018}}

@article{envbench,   title={Envbench: A benchmark for automated environment setup},
  author={Eliseeva, Aleksandra and Kovrigin, Alexander and Kholkin, Ilia and Bogomolov, Egor and Zharov, Yaroslav},
  journal={arXiv preprint arXiv:2503.14443},
  year={2025}}

@article{repo2run, title={Repo2run: Automated building executable environment for code repository at scale}, author={Hu, Ruida and Peng, Chao and Xu, Junjielong and Gao, Cuiyun}, journal={Advances in Neural Information Processing Systems}, volume={38}, pages={32679--32718}, year={2026}}

@article{scireplicate, title={Scireplicate-bench: Benchmarking llms in agent-driven algorithmic reproduction from research papers}, author={Xiang, Yanzheng and Yan, Hanqi and Ouyang, Shuyin and Gui, Lin and He, Yulan}, journal={arXiv preprint arXiv:2504.00255}, year={2025}}

@article{paperbench, title={PaperBench: Evaluating AI's Ability to Replicate AI Research}, author={Starace, Giulio and Jaffe, Oliver and Sherburn, Dane and Aung, James and Chan, Jun Shern and Maksin, Leon and Dias, Rachel and Mays, Evan and Kinsella, Benjamin and Thompson, Wyatt and others}, journal={arXiv preprint arXiv:2504.01848}, year={2025}}

@article{repomaster, title={Repomaster: Autonomous exploration and understanding of github repositories for complex task solving}, author={Wang, Huacan and Ni, Ziyi and Zhang, Shuo and Lu, Shuo and Hu, Sen and He, Ziyang and Hu, Chen and Lin, Jiaye and Guo, Yifu and Du, Yuntao and others}, journal={Advances in Neural Information Processing Systems}, volume={38}, pages={106320--106359}, year={2026}}

@inproceedings{gittaskbench, title={Gittaskbench: A benchmark for code agents solving real-world tasks through code repository leveraging}, author={Ni, Ziyi and Wang, Huacan and Zhang, Shuo and Lu, Shuo and He, Ziyang and Tang, Zhenheng and Hu, Sen and Li, Bo and Hu, Chen and Jiao, Binxing and others}, booktitle={Proceedings of the AAAI Conference on Artificial Intelligence}, volume={40}, number={38}, pages={32564--32572}, year={2026}}

@inproceedings{r2e, title={R2e: Turning any github repository into a programming agent environment}, author={Jain, Naman and Shetty, Manish and Zhang, Tianjun and Han, King and Sen, Koushik and Stoica, Ion}, booktitle={Forty-first International Conference on Machine Learning}, year={2024}}

@inproceedings{creator, title={Creator: Tool creation for disentangling abstract and concrete reasoning of large language models}, author={Qian, Cheng and Han, Chi and Fung, Yi and Qin, Yujia and Liu, Zhiyuan and Ji, Heng}, booktitle={Findings of the Association for Computational Linguistics: EMNLP 2023}, pages={6922--6939}, year={2023}}

@inproceedings{bfcl, title={The berkeley function calling leaderboard (bfcl): From tool use to agentic evaluation of large language models}, author={Patil, Shishir G and Mao, Huanzhi and Yan, Fanjia and Ji, Charlie Cheng-Jie and Suresh, Vishnu and Stoica, Ion and Gonzalez, Joseph E}, booktitle={Forty-second International Conference on Machine Learning}, year={2025}}

@inproceedings{metagpt, title={MetaGPT: Meta programming for a multi-agent collaborative framework}, author={Hong, Sirui and Zhuge, Mingchen and Chen, Jonathan and Zheng, Xiawu and Cheng, Yuheng and Wang, Jinlin and Zhang, Ceyao and Yau, Steven and Lin, Zijuan and Zhou, Liyang and others}, booktitle={International Conference on Learning Representations}, volume={2024}, pages={23247--23275}, year={2024}}

@article{autogen, title={Autogen: Enabling next-gen llm applications via multi-agent conversation}, author={Wu, Qingyun and Bansal, Gagan and Zhang, Jieyu and Wu, Yiran and Li, Beibin and Zhu, Erkang and Jiang, Li and Zhang, Xiaoyun and Zhang, Shaokun and Liu, Jiale and others}, journal={arXiv preprint arXiv:2308.08155}, year={2023}}

@inproceedings{agentverse, title={Agentverse: Facilitating multi-agent collaboration and exploring emergent behaviors}, author={Chen, Weize and Su, Yusheng and Zuo, Jingwei and Yang, Cheng and Yuan, Chenfei and Chan, Chi-Min and Yu, Heyang and Lu, Yaxi and Hung, Yi-Hsin and Qian, Chen and others}, booktitle={International Conference on Learning Representations}, volume={2024}, pages={20094--20136}, year={2024}}

@article{mad, title={Improving factuality and reasoning in language models through multiagent debate}, author={Du, Yilun and Li, Shuang and Torralba, Antonio and Tenenbaum, Joshua B and Mordatch, Igor}, journal={arXiv preprint arXiv:2305.14325}, year={2023}}

@article{madstudy, title={Can LLM Agents Really Debate? A Controlled Study of Multi-Agent Debate in Logical Reasoning}, author={Wu, Haolun and Li, Zhenkun and Li, Lingyao}, journal={arXiv preprint arXiv:2511.07784}, year={2025}}

@article{principalagent, title={Multi-Agent Systems Should be Treated as Principal-Agent Problems}, author={Rauba, Paulius and Cepenas, Simonas and van der Schaar, Mihaela}, journal={arXiv preprint arXiv:2601.23211}, year={2026}}

@article{knowno, title={Robots that ask for help: Uncertainty alignment for large language model planners}, author={Ren, Allen Z and Dixit, Anushri and Bodrova, Alexandra and Singh, Sumeet and Tu, Stephen and Brown, Noah and Xu, Peng and Takayama, Leila and Xia, Fei and Varley, Jake and others}, journal={arXiv preprint arXiv:2307.01928}, year={2023}}

@article{sagallm, title={SagaLLM: context management, validation, and transaction guarantees for multi-agent LLM planning}, author={Chang, Edward Y and Geng, Longling}, journal={arXiv preprint arXiv:2503.11951}, year={2025}}

@article{dover, title={Dover: Intervention-driven auto debugging for llm multi-agent systems}, author={Ma, Ming and Zhang, Jue and Yang, Fangkai and Kang, Yu and Lin, Qingwei and Rajmohan, Saravan and Zhang, Dongmei}, journal={arXiv preprint arXiv:2512.06749}, year={2025}}

@article{causalflow, title={CausalFlow: Causal Attribution and Counterfactual Repair for LLM Agent Failures}, author={Bonagiri, Akash and Borkar, Devang and Anderias, Gerard Janno and Rafatirad, Setareh and Homayoun, Houman}, journal={arXiv preprint arXiv:2605.25338}, year={2026}}

@article{tms, title={A truth maintenance system}, author={Doyle, Jon}, journal={Artificial intelligence}, volume={12}, number={3}, pages={231--272}, year={1979}, publisher={Elsevier}}

@article{eken2025multivocal,
  title={A multivocal review of MLOps practices, challenges and open issues},
  author={Eken, Beyza and Pallewatta, Samodha and Tran, Nguyen and Tosun, Ayse and Babar, Muhammad Ali},
  journal={ACM Computing Surveys},
  volume={58},
  number={2},
  pages={1--35},
  year={2025},
  publisher={ACM New York, NY}
}

@article{paleyes2022challenges,
  title={Challenges in deploying machine learning: a survey of case studies},
  author={Paleyes, Andrei and Urma, Raoul-Gabriel and Lawrence, Neil D},
  journal={ACM computing surveys},
  volume={55},
  number={6},
  pages={1--29},
  year={2022},
  publisher={ACM New York, NY}
}

@article{artifactdeploy,
  title={DeployBench: Benchmarking LLM Agents for Research Artifact Deployment},
  author={Wang, Yuanli and Qian, Yaoyao and Zhang, Yue and Zhou, Hanhan and Huang, Jindan and Fu, Tianfu and Mang, Qiuyang and Mao, Huanzhi and Chai, Wenhao and Fan, Wendong and others},
  journal={arXiv preprint arXiv:2606.05238},
  year={2026}
}


\end{document}